\documentclass[english]{article}
\usepackage[T1]{fontenc}
\usepackage[latin9]{inputenc}
\usepackage{amstext}
\usepackage{graphicx}
\usepackage{babel}
\makeatletter
\newcommand{\lyxaddress}[1]{
\par {\raggedright #1
\vspace{1.4em}
\noindent\par}
}

\makeatother

\begin{document}

\title{MAGNETIC MEASUREMENTS AND KINETIC ENERGY OF THE SUPERCONDUCTING CONDENSATE
IN SmBa$_{2}$Cu$_{3}$O$_{7-\delta}$}

\author{J.P. Pe\~na$^{1,2}$, D.B. Mart\'inez$^{1}$, C.A. Parra Vargas$^{1}$,
A.G. Cunha$^{\text{3}}$,\\ J.L. Pimentel Jr.$^{2}$, P. Pureur$^{2}$ }

\maketitle

\lyxaddress{$^{1}$ Grupo de F\'isica de Materiales, Universidad Pedag\'ogica y Tecnol\'ogica
de Colombia, Av. Central del Norte, Tunja, Colombia. }

\lyxaddress{$^{2}$ Instituto de F\'isica, Universidade Federal do Rio Grande do
Sul, Av. Bento Gon\c{c}alves 9500, C.P. 15051, 91501-970, Porto Alegre,
RS, Brazil.}

\lyxaddress{$^{\text{3}}$Departamento de F\'isica, Universidade Federal do Esp\'irito
Santo, Av. Fernando Ferrari 514, Campus Goiabeiras, 29075-910, Vit\'oria,
ES, Brazil.}

\begin{abstract}
We report in-field kinetic energy results in the temperature region
closely below the transition temperature of two differently prepared
polycrystalline samples of the superconducting cuprate SmBa$_{\text{2}}$Cu$_{\text{3}}$O$_{7-\delta}$.
The kinetic energy was determined from magnetization measurements
performed above the irreversibility line defined by the splitting
between the curves obtained according the ZFC and FC prescriptions.
The results are analyzed in the intermediate field regime where the London approximation can 
be used for describing the magnetization. 
From the analysis, estimations were carried out for the penetration depth 
and the upper cri\-ti\-cal field of the studied samples.The difference between the kinectic energy magnitudes for the
two studied samples is ascribed to effects from granularity. 
\end{abstract}

\section{Introduction}

The experimental study of the kinetic energy of the charge carriers
in the high temperature cuprate superconductors (HTCS) become an important
subject in view  of the new theoretical approaches predicting
that, in the absence of an applied magnetic field, pairing in these materials results from a decrease of 
the kinetic energy
term upon condensation {[}1,2{]}. This prediction is opposite to that
of the BCS theory, where the kinetic energy of the carriers increases
when the system enters the superconducting state {[}3{]}. Experimentally,
the situation is less clear because of the relatively small change
of the kinetic energy in the condensate with respect to that of the
normal state. Optical conductivity measurements in optimally doped
and underdoped Bi$_{2}$Sr$_{2}$CaCu$_{2}$O$_{8+d}$ (Bi-2212) suggests
that the observed spectral weight transfer that occurs below the superconducting
transition is in agreement with the unconventional scenario of decreasing
kinetic energy {[}4{]}. On the other hand, when this cuprate is slightly
overdoped, results are rather in accordance with the BCS predictions
{[}5{]}.  

A simpler situation occurs in a type II superconductor when a magnetic
field is applied. In that case, it is expected that the kinetic energy of the condensate
always  increase because of the flux expulsion and vortex formation
{[}6{]}. Useful informations may be obtained on the order parameter and other basic properties of type II superconductors
from studies of this energy term {[}7{]}. In the case discussed here, we are interested in 
the kinetic energy associated to the currents around the vortices generated by the action of an applied
 external magnetic field.

Theoretically, it was demonstrated
from applying the virial theorem in the framework of the Ginzburg-Landau
(G-L) theory {[}8{]} that the average kinetic energy density of a
large $\kappa$ superconductor may be written as {[}6,7{]}
\begin{equation}
E_{K}=-\vec{M}\cdot\vec{B}\:,\end{equation}
where $\vec{M}$ is the equilibrium magnetization and $\vec{B}$ is
the magnetic induction. Inside a superconducting sample, the induction
(in the SI system) may be written as $\vec{B}=\mu_{0}\vec{H}+\mu_{0}(1+\eta)\vec{M}$,
where $\mu_{0}$ is the vacuum permeability, $\vec{H}$ is the applied
field and $\eta$ is the sample dependent geometric factor related
to the dipolar field. Then, for obtaining the kinetic energy density
from magnetization measurements, it is useful to re-write equation
(1) as

\begin{equation}
E_{K}=-\mu_{0}\vec{M}\cdot\vec{H}-\mu_{0}(1+\eta)\, M^{2}.\end{equation}

The in-field kinectic energy density was determined for some low and
high $T_{c}$ superconductors in the references {[}6{]},{[}9{]} and {[}10{]}. The results reported for
Nb {[}9{]} and for a Pb-In alloy {[}10{]} are in agreement with the
expectations derived from the Abrikosov treatment of the G-L theory,
as well as with general BCS predictions. On the other hand, in cuprates
as optimally doped and underdoped YBa$_{2}$Cu$_{3}$O$_{x}$ (YBCO)
{[}9{]}, optimally doped Bi-2212 {[}9{]} and La$_{1.9}$Sr$_{0.1}$CuO$_{4}$
{[}10{]} an appreciable amount of the in-field kinetic energy subsists
up to temperatures well beyond $T_{c}$. Authors in references {[}9{]}
and {[}10{]} identify this behavior to a pseudogap effect, although
the influence of strong thermal fluctuations cannot be ruled out as
an alternative explanation.

In order to study the differences induced by the microscopic morphology in the kinetic energy, here
 we report magnetization measurements and in-field
kinetic energy density estimations in two different polycrystalline samples
of SmBa$_{2}$Cu$_{3}$O$_{7-\delta}$ (Sm-123). The experiments were
performed in several applied fields. However, the extraction of $E_{K}(T,H)$
is restricted to the temperature range near $T_{c}$ , where the equilibrium
magnetization can be unambiguously obtained from coincident and reproducible
ZFC and FC measurements. Results are compared to those obtained in
YBCO and Bi-2212 single crystals {[}9,10{]}. The magnitude of the kinetic energy of our 
polycrystalline samples is considerably smaller than that of single crystal samples.

\section{Experimental}

Two samples of polycrystalline SmBa$_{2}$Cu$_{3}$O$_{7-\delta}$
(Sm-123) were independently produced using the solid state reaction
method. The employed procedures are described below. The precursor compounds Sm$_{2}$O$_{3}$
(purity 99.99\%), BaCO$_{3}$ (purity 99.8\%) and CuO (purity 99,995\%) were used for the preparation
of the samples.
 For the sample labeled as Sm-I, the presursors were mixed and macerated during two hours in an agate mortar, pressed
at 3 tons into a cilindrical shape and submitted to a calcination
process at 850 $^{0}C$ during 15 hours. After furnace cooling to
room temperature, the resulting pellet was finely powdered again,
pressed and subjected to two sintering processes at 870 $^{0}C$ and
890 $^{0}C$ for 45 hours each one. The cylindrical sample was then oxygenated by exposure
 to an oxygen atmosphere while decreasing the temperature in the range 750-250 $^{0}C$
 at rate of 12.5 $^{0}C/$h, then it was furnace cooled to room temperature.
The preparation of sample Sm-II was also done in three steps but using
different equipments and thermal processes. In the calcination
process, the macerated and pressed precursors were heated up to 850
$^{0}C$ and kept at this temperature during 48 hours. For the sintering
step, the sample was heated up to 900 $^{0}C$ at 150 $^{0}C/$h,
kept at this temperature for 0.1 h, then heated up to 1040
$^{0}C$ at 60 $^{0}C/$h. A 24 hours annealing was carried out at
this temperature before cooling the sample at -60 $^{0}C/$h to 900
$^{0}C$ where it was maintained for 0.1 h. Subsequently, the
sample was cooled down to room temperature at 150 $^{0}C/$h. In the
final oxygenation process, the sample Sm-II was heated again to 500
$^{0}C$ and kept at this temperature during 0.1 h. Then, it
was cooled to 350 $^{0}C$ and kept at this temperature for 3 days.
After that time, the sample was furnace cooled to room temperature.

 Electrical resistivity measurements
were performed in both samples. Transition temperatures determined
from the maximum of the temperature derivative of the resistivity
are $T_{c}=92.7$ K for sample Sm-I and $T_{c}=89.9$ K for sample
Sm-II. The width of the resistive transition is around 3 K for both
samples.

Zero field cooling (ZFC) and field cooling (FC) magnetization measurements
in several applied fields ranging from 1 mT up to 5 T were performed
using a XL5-MPMS SQUID magnetometer manufactured by Quantum Design
Inc. Results were corrected for the demagnetization effects. The geometrical
factors were estimated by approximating the samples's shape to an
ellipsoid and making use of the calculations in reference {[}11{]}.
The porosity of the ceramic samples was also taken into account by
comparing their measured densities and the theoretically expected
value.

\section{Results}

\subsection{Reversible Magnetization and Fluctuation Effects}

Figure 1 shows representative magnetization versus temperature results
in a temperature range encompassing the superconducting transition
for the two studied samples. Panel (a) shows results for sample Sm-I
and in panel (b) are plotted data for sample Sm-II. Experiments were
performed according the ZFC and FC prescriptions. The irreversibility
temperature \emph{T$_{irr}$} denotes the point below which the pinning
effects become important. We carried out our analyses in the temperature
regime above \emph{T}$_{irr}$, where the experimental data describe
the equilibrium magnetization without ambiguity. The extension
of the magnetically reversible regime in high applied fields is smaller for the sample Sm-II than for the sample Sm-I.

\begin{figure}[h]
\centering
\includegraphics[scale=0.34]{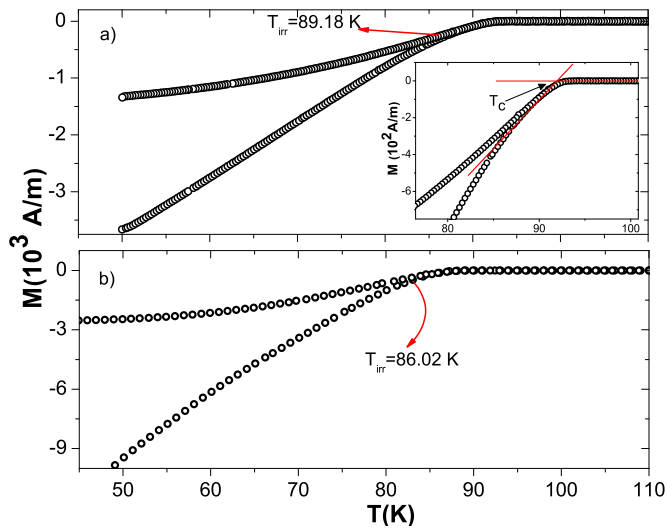}
\caption{Zero field cooled (ZFC) and field cooled (FC) magnetization as functions
of the temperature for polycrystalline SmBa$_{2}$Cu$_{3}$O$_{7-\delta}$
(Sm-123) measured at \emph{$\mu_{0}$H} = 0.05 T. Panel (a) shows
results for sample Sm-I (see text), the inset illustrates  the criterion
used to define the in-field critical temperatures. Panel (b) shows results for sample Sm-II.
The irreversibility temperature \emph{T}$_{irr}$ is signaled and
denotes the splitting of the ZFC and FC curves. Corrections for the
demagnetizing field effects are taken into account.}
\end{figure}

Figures 2(a) and 2(b) collects the characteristic temperatures $T_c$ and $T_irr$ in 
the presence of several applied fields for samples Sm-I and Sm-II, respectively. The area delimited
between the curves $T_{c}(H)$ and $T_{irr}(H)$ defines the reversible
region where effects of the pinning energy are negligible. The
field-dependent critical temperatures are determined by the intersection
between the linearly extrapolated magnetizations in the normal and
superconducting phases, as illustrated in the inset of Fig. 1(a). The
irreversibility temperatures are determined by subtracting the ZFC
magnetization from the FC one and applying the criterion used in
reference {[}12{]}: the temperature where the difference $M_{FC}(T,H)-M_{ZFC}(T,H)$
deviates from the zero baseline defined by the high temperature data
is assigned to $T_{irr}(H)$. 

\begin{figure}[h]
\centering
\includegraphics[scale=0.28]{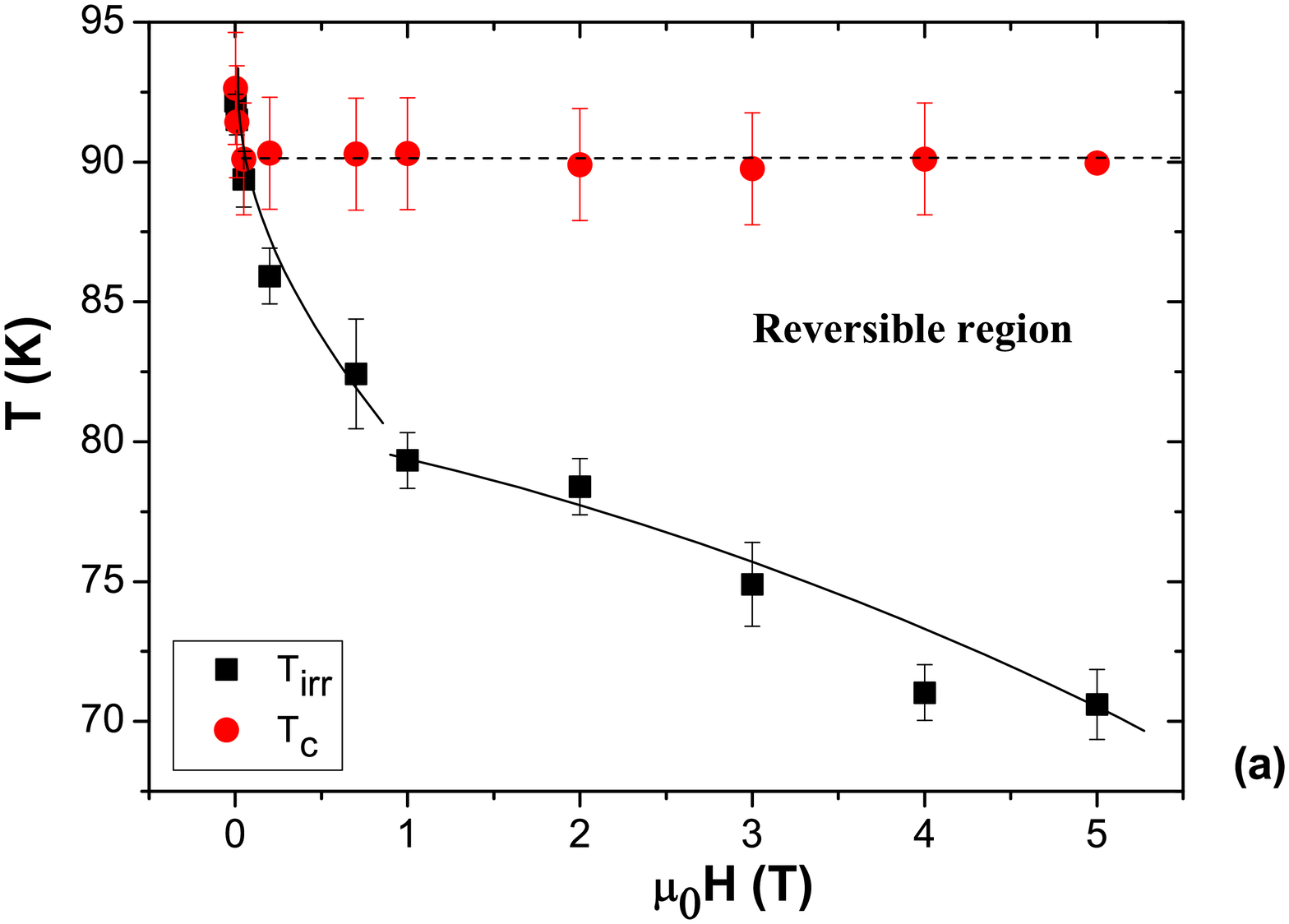}
\includegraphics[scale=0.26]{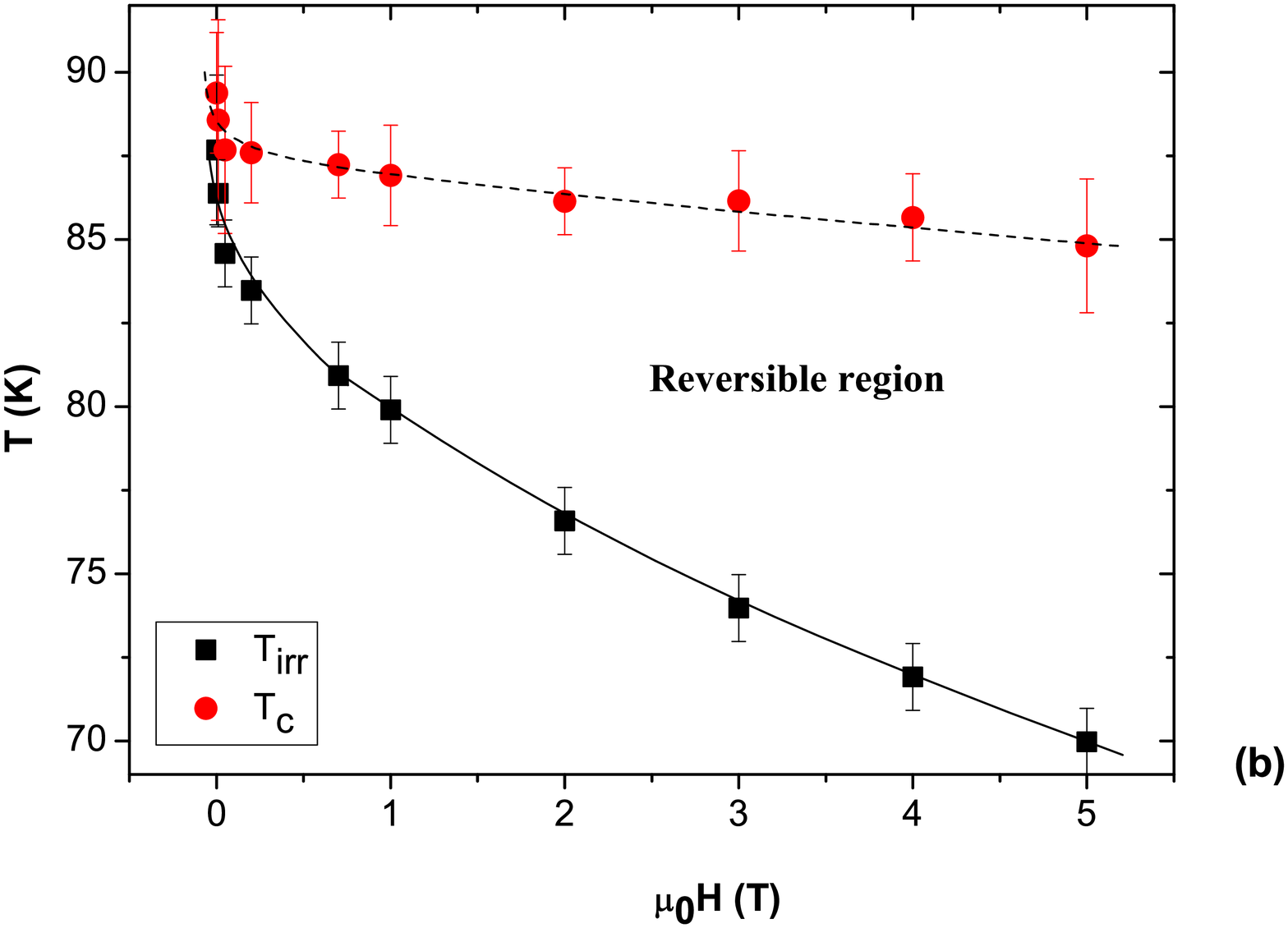}
\caption{Critical temperatures and irreversibility temperatures as functions
of the applied field for the samples Sm-I (a) e Sm-II (b). The continuous and dashed lines are guides for the eye.}
\end{figure}

The observed behaviour of $T_{irr}(H)$ on the two samples
is typical of polycrystalline samples of HTCS {[}13{]} and mimics
the observations in spin glass systems. For the sample Sm-I, in the low field and high
temperature limit, the irreversibility temperatures decrease as a
function of the applied field as $[T_{irr}(0)-T_{irr}(H)]\sim H^{2/3}$,
in analogy to the de Almeida-Thouless line {[}14{]}. As shown in Fig.
2(a), near $\mu_{0}H=0.5$ T, a crossover is observed in $T_{irr}(H)$
to a field dependence with inverse curvature that may be approximately
fitted to a Gabay-Toulouse-like line {[}15{]} given by $[T_{irr}(0)-T_{irr}(H)]\sim H^{2}$.
The behaviour observed in Fig. 2(a) was interpreted as the vortex glass
analog of the crossover from the high-field Gabay-Toulouse transition
to the low field de Almeida-Thouless instability observed in Heisenberg
vector spin glasses {[}16{]}. 
As shown by Fig. 2(b) for sample Sm-II, a de Almeida-Thouless-like line describes the irreveribility line 
in the whole field range in this case. The results in Fig. 2 suggest that the effects from granularity are 
stronger in sample Sm-I.

\subsection{Kinetic Energy}

We assume that the G-L theory describes the equilibrium
magnetization adequately in most of the reversible superconducting
regime of our samples. Then, using the recipe of Doria el al. {[}6{]},
in Figs. 3 (a) and (b) we plot  the field induced kinectic energy  given
by equation (2) as a function of temperature  for samples Sm-I
and Sm-II, respectively. Results for magnetic fields ranging from 0.2 T up to
5 T are shown and the data are restricted to the temperature regime
where the magnetization is reversible. The applied field range in our
investigation enlarges significantly the regime studied in Ref. {[}9{]}
for YBCO. Here, the kinetic energy density approaches linearly to the temperature 
axis and vanishes for $T\approx T_{c}(H)$. We do not observe a significant kinetic energy
 contribution above $T_{c}$, contrasting with the findings reported in Ref. {[}9{]}.
On the other hand, $E_{K}$ in Sm-123 and YBCO are qualitatively similar
in the field and temperature range where the comparison is possible, even
though results in YBCO were obtained for fields applied perpendicular
to the Cu-O$_{\text{2}}$ atomic layers of a single crystal sample
{[}9{]}. 

Although the similar field and temperature behavior shown by results
for samples Sm-I and Sm-II in Fig. 3, these data differ quantitatively.
For a given field and temperature, $E_{K}$ is significantly larger
for sample Sm-II.
We attribute this difference to polycrystalinity.
Indeed, the different routes employed to prepare our samples likely 
lead to distinct granular microstrustures and granularity is known to underly 
the magnetic behavior of polycrystalline samples of the HTCS. 

\begin{figure}
\centering
\includegraphics[scale=0.25]{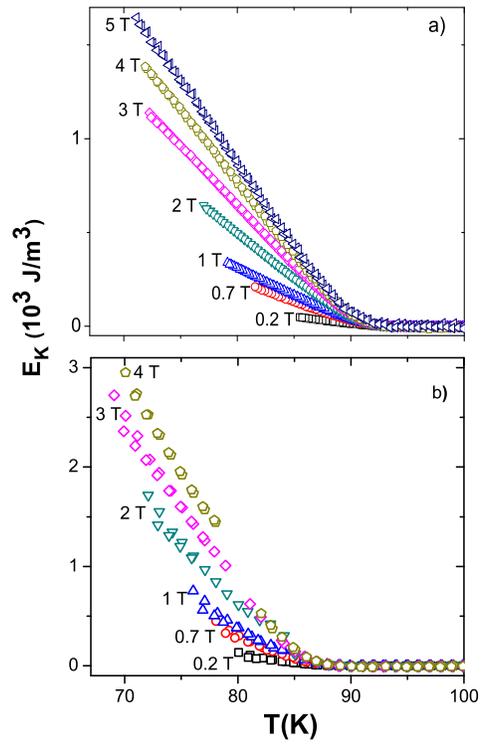}
\caption{Kinetic energy density as a function of the temperature for samples
(a) Sm-I and (b) Sm-II in the quoted applied fields.}
\end{figure}

 In the intermediate field regime, $H_{c1}\ll H\ll H_{c2}$, 
  the interaction among the vortices is weak {[}3{]}. 
In this field region, the Cooper pair density may be considered uniform 
inside the sample excepting the vortex positions. In the case of extreme type II
 superconductors ($\kappa>>1)$ as the HTCS, the variation of the order parameter in 
the vortex positions may be described by delta functions. Under such conditions the 
magnetization can be calculated with the London approximation to the G-L theory. In this case,
 the magnetization in SI units is {[}17{]}: 

\begin{equation}
 M(H)=-\frac{\phi_0}{8\mu_0\pi\lambda^2}\ln\left(\frac{\beta_LH_{c2}}{H}\right),
\end{equation}
where $\phi_0$ is the quantum magnetic flux, $\mu_0$ is the vacuum permeability, $\lambda$ is the penetration length and
$\beta_L$ is a constant of order unity. 

The substitution of the magnetization given by equation (3) in equation (2) yields:
\begin{equation}
  \frac{E_{K}(\mu_0H)}{\mu_0H}=\frac{\phi_0}{8\pi\lambda^2\mu_0}\ln\frac{\beta_L\mu_0H_{c2}}{\mu_0H}-
\left(\frac{\phi_0}{8\pi\lambda^2}\right)^2
\frac{1}{\mu_0^2H}\left(\ln\frac{\beta_L\mu_0H_{c2}}{\mu_0H}\right)^2.
\end{equation}
Fittings of the experimental data to equation (4) are displayed in Fig. 4 for samples Sm-I and Sm-II.
 From these fittings we estimate the penetration lengths and the upper critical field in the studied field
 and temperature ranges. The values obtained for $\lambda (T)$ and $H_{c2}(T)$ were corrected according the
 Hao and Clem model {[}18{]} and are shown in figs. 5 and 6, respectively.  

 \begin{figure}[h]
 \centering
 \includegraphics[scale=0.40]{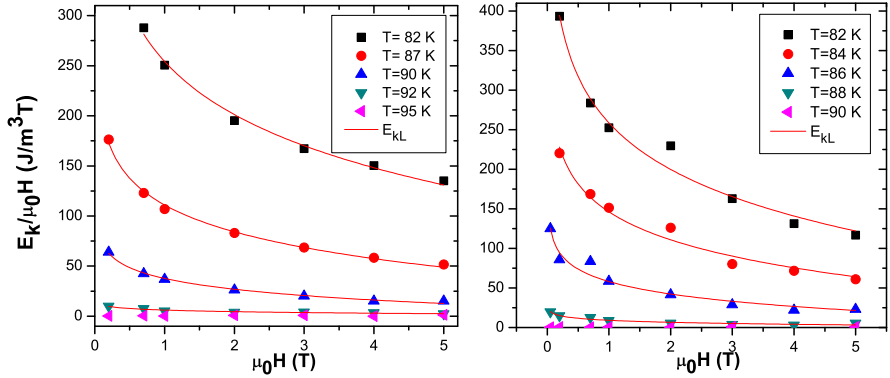}
 \caption{Kinetic energy per unit of field for the samples Sm-I (left) and Sm-II (rigth). The solid lines 
are fits of eq. (4) to the experimental data.} 
 \end{figure}

According to the mean-field theory, $\lambda(T)=\frac{1}{\sqrt{2}}\lambda(0)t^{-1/2}$
 and $B_{c2}(T)=1,83B_{c2}(0)t$ where $t=\frac{T_c-T}{T_c}$ is the reduced temperature {[}19{]}. In fig. 5, 
the values for $\lambda(T)$ for samples 
Sm-I and Sm-II are plotted as functions of $1/\sqrt{t}$. The fitted straigth lines allow us to estimate 
$\lambda(0)=516\pm12$ nm for sample Sm-I and $\lambda(0)=363\pm7$ nm
for sample Sm-II. These values are within the expected range, since the estimated penetration lenghts are
 polycrystalline averages enhanced by granularity effects {[}20{]}. The larger $\lambda(0)$ found
 in sample Sm-I also suggests that the influence of granularity is stronger in this sample. 

 \begin{figure}[h]
 \centering
 \includegraphics[scale=0.28]{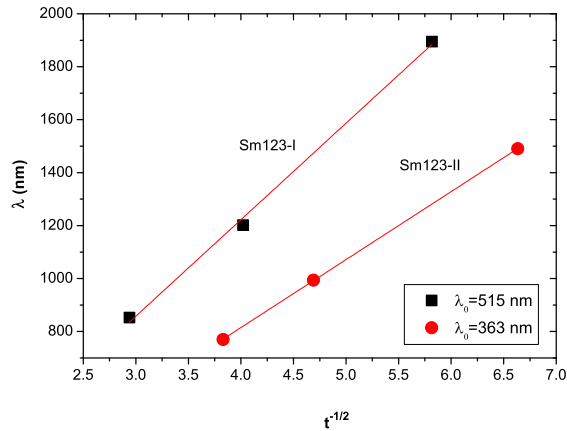}
 \caption{Penetration lengths for samples Sm-I and Sm-II deduced from fits of the data in fig. 4 to equation (4). 
Plots are made as functions of $1/\sqrt{t}$.} 
 \end {figure}

In fig. 6, the estimations for $H_{c2}$ extracted from fittings in fig.4 are plotted as a function 
of the reduced temperature. Data for both samples align to a single straight line. From the slope 
of line we deduce that $B_{c2}(0)=130$ T for Sm-123. This estimation is in good aggreement 
with previous determinations of polycrystalline averages of the upper critical induced field in Y-123-type
 superconductors {[}19{]}. 

 \begin{figure}[h]
 \centering
 \includegraphics[scale=0.30]{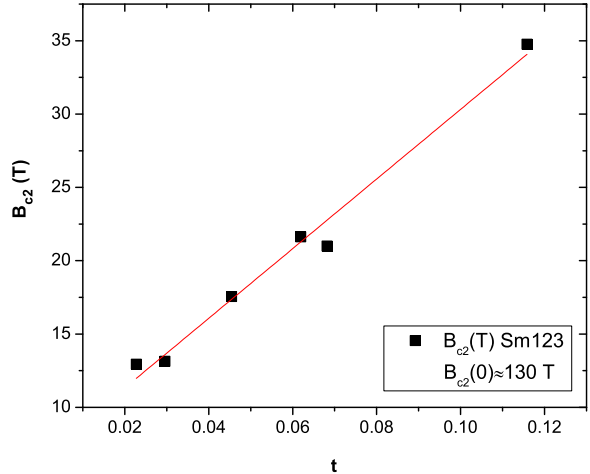}
 \caption{Upper critical field for samples Sm-I and Sm-II as deduced from the results in fig. 4.
 Data points are plotted as a function of the reduced temperature. A single straigth line fits the 
data for both samples.}
 \end {figure}

Since granularity leads the polycrystalline HTCS systems to behave as a 
superconducting glassy medium {[}21{]}, the in-field kinetic energy in our Sm-123 samples 
may not be enterely accounted  by the London approximation 
to the Ginzburg-Landau theory. So, one might consider that a complete description of the kinetic energy
 in granular HTCS should include the contributions of Josephson vortices {[}22{]} 
 and intergrain chiralities {[}23{]}. However, as a first approximation, equation (4) furnishes a fairly 
good descrition of the data, showing that the intragrain vortices are indeed the dominant contribution to 
the condensate kinetic energy in Sm-123 in the investigated field-temperature range.

\section{Conclusions}

We studied experimentally the reversible magnetization and the field-induced kinetic energy density in 
two polycrystalline samples of the SmBa$_{\text{2}}$Cu$_{\text{3}}$O$_{\text{7-\ensuremath{\delta}}}$ 
cuprate superconductor. Analysis were restricted to the temperature range nearly below $T_{c}$, where
 the ZFC and FC magnetization are coincident. Fields up to 5 T were applied. 

The theory proposed by M. Doria and co-workers {[}6,7{]} which is based on the application of 
the virial theorem to the Ginzburg-Landau free energy was used to describe the kinetic energy 
density in our samples. The magnitude of the kinetic energy was found larger for sample Sm-II than 
for sample Sm-I. We attributed this difference to the enhanced granularity effects in sample Sm-I.
 The identified contribution to the kinetic energy in both samples were attributed to intragrain vortices. 
Results could be interpreted by assuming the validity of London approximation to the Gizburg-Landau theory.
 The penetration length and the upper critical field were calculated for both samples. The obtained values 
are similar to those found in previous studies. The penetration lenght estimations corroborate the enhanced 
effects of granularity in sample Sm-I. 

As a final conclusion, our results and analysis showed that the study of the in-field kinetic 
energy may improve the usual description of the magnetic response of superconductors based solely 
on magnetization results in the temperature range approaching $T_c$, where the pinning effects are inexistent.

\paragraph*{Acknowledgments}

We thank Dr. Mauro Doria for enlightening discussions on the in-field
kinetic energy of a superconducting condensate. This work was partially
financed by the Brazilian agencies FAPERGS and CNPq under the grant
PRONEX 10/0009-2.

\end{document}